# Finite Element Modeling of Micro-cantilevers Used as Chemical Sensors

G. Louarn and S. Cuenot
*Institut des Matériaux Jean Rouxel*
*University of Nantes, France*

During the last twenty years, the spectacular advances in micro-mechanical systems, and more generally in micro-electro-mechanical systems (MEMS), have enabled the emergence of an innovative family of biological and chemical sensors. The functionality of such sensors involved the transduction of a mechanical energy, which comes from the deformations of the micro-machined components. Among the different geometrical shapes of these components, clamped-free beams also called cantilevers represent the simplest MEMS. Although cantilevers play a rule of basic building block for complex MEMS devices, they are nowadays widely used as force sensor probes in atomic force microscopy (AFM).

The recent developments of AFM have enabled the mass-fabrication of micrometer-sized cantilevers with various geometrical shapes and different materials (silicon, silicon nitride…). Indeed, beam cantilevers but also V-shaped cantilevers are commercially available with different dimensions. Recently, this variety of AFM cantilevers is become the base of a novel class of highly sensitive sensors operating essentially for chemical and biological detections (Raiteri et al., 2001 ; Moulin et al., 2003 ; Raiteri et al., 2001 ; Lavrik et al., 2004). These micro-cantilevers, which can be extremely versatile sensors, present many advantages such as a relatively low cost of production and a reduced size of the active area (typically $10^{-6}$ cm²). The detection principle of such sensors is based on the measurement of the cantilever deflection change or of the resonance frequency change of the cantilever (Lavrik et al., 2004). These changes are induced by the adsorption of chemical species on the functionalized surface of a micro-cantilever (Chen, 1995).

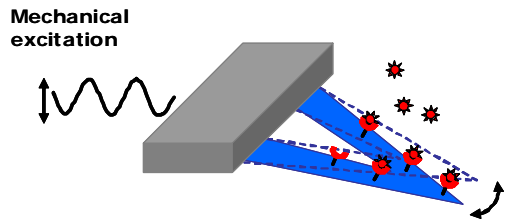

Such adsorption may come from the interactions between specific molecules present in analyte and the sensitive coating of the cantilever. The measurement of the resonance frequency shift offers the interesting advantage to be relatively insensitive to interference from external factors such as thermal drift. Therefore, this method seems particularly well-suited for use in various environments such as gaseous or vacuum ones, and to measure the cantilever response upon exposure to optical radiation and to chemical vapours. Several applications of these AFM micro-cantilevers as sensors were developed during the ten last years. According to the realized coating, the chemically modified surfaces of cantilevers can be used to produce transducers that are activated for specific analytes. For instance, cantilevers coated with a thin gold film become ideal sensors to achieve ultrasensitive detection of mercury vapour (Thundat et al., 1995). A metallic coating allows the calorimetric detection of chemical reactions with picojoule sensitivity (Gimzewski et al., 1994), while a gelatine coating is sensitive to the changes in relative humidity (Wachter & Thundat et al., 1995). Others detection can also be achieved such as pH-variation (Ji et al., 2001 ; Bashir et al., 2002) or optical radiation by coating the cantilevers with ultraviolet cross-linking polymers (Thundat, 1995).

In order to predict the sensitivity of these versatile sensors, it is crucial to precisely calculate the resonant frequencies of these micro-cantilevers. The resonance frequencies can be only calculated analytically for cantilevers with simple geometries like straight beams. In the other cases, for multilayer structures or more complex shapes, where the cross-section is not constant along the cantilever length (case of "V-shaped" micro-cantilevers), the resonant frequencies can not be analytically calculated. Among the different approaches allowing to calculate them, the finite element modeling (FEM) seems to be a particularly interesting approach. In this chapter, FEM is applied to "V-shaped" silicon micro-cantilevers for predicting their resonant frequencies and the sensitivity of these cantilevers employed as chemical sensors.

First, in order to validate the accuracy of the FEM approach, we carried out a comparison between analytical, experimental and FEM-computed values of the resonant frequencies for homogenous rectangular shaped micro-cantilevers. Then, we performed a modeling of silicon beams coated with a thin sensitive layer (50 nm of Gold). To precisely calculate the resonant frequencies of these multilayer-cantilevers, the influence of the mesh parameters on the calculated frequencies was strongly investigated.

Second, the sensitivity of different "V-shaped" silicon cantilevers was estimated, as a function of their geometrical dimensions and of their mechanical parameters (Young modulus, density). The resonant frequencies of uncoated cantilevers were calculated and compared with the values experimentally determined. Then, a similar approach could be employed to predict the sensitivities of such cantilevers recovered with a sensitive layer.
This chapter clearly shows that the use of finite element modeling allows the computation of the resonant frequencies of micro-sized cantilevers having a complex shape. Moreover, in the case of applications as sensors based on the resonant frequency measurement, this approach offers the ability to predict the sensitivity of multilayer-cantilevers. In this way, by simple calculation, the shape and the geometrical dimensions of these cantilevers can be optimized for obtaining the best sensitivities and detection limits.

## 2. Natural frequencies and normal modes of vibration

When a linear mechanical system undergoes free vibrations, i.e. without being subjected to any external periodic load, its behaviour is characterized by its so-called "eigen modes". An eigen mode is a global synchronous type of motion where the various degrees of freedom exhibit all the same time behaviour (with different amplitudes), viz. harmonic motion at a frequency which is characteristic of the mode. An n degrees of freedom system has n eigen modes, hence n characteristic eigen frequencies, also called "natural" frequencies.

When the system undergoes forced harmonic vibration, its behaviour is characterized by the frequency response relating the amplitude (and phase) of the steady-state harmonic response to that of the excitation. The main feature of this response is a series of peaks appearing at forcing frequencies close to the natural frequencies. These large response amplitudes (theoretically infinite if no damping is present) relate to the so-called "resonant" behaviour of the system – generally to be avoided in practice in most engineering application.

When the system is undamped, the (infinite) resonance peaks appear at forcing frequencies, which are identical to the natural frequencies of the system. From a physical point of view, since the system is excited at a frequency at which it would tend to freely oscillate, its vibration amplitude increases steadily, tending towards infinity.

In the case of a damped system, there are various definitions of the related "resonant" frequencies. One of these is the forcing frequency (close to a natural frequency) at which the maximum of a peak is reached. However, at a forcing frequency equal to a natural frequency, the phase shift between the response and the excitation is always an odd multiple of 90°, and this is generally chosen as the most reliable and accurate definition of resonance.

We see therefore that "natural" and "resonance" frequencies are two names for the same notion, albeit related to two different situations ("free" and "forced" vibrations). In the following, we shall use the term "natural" frequencies when analyzing free vibrations, and "resonance" frequencies when dealing with forced vibrations.

For systems having distributed parameters, partial differential equations are obtained. Unfortunately, exact solutions of theses equations are possible only for few configurations. For our V-shaped cantilevers, other approaches must be employed to solve the equations. Research

**Procédure de mesure de fréquences par AFM**
La première partie de ce travail consiste à mesurer à l'aide d'un AFM les fréquences de résonance de différents leviers. Cette mesure s'effectue dans le mode de fonctionnement de l'AFM dit TappingMode™. L'appareil utilisé est un Multimode AFM VEECO® Nanoscope IIIa. Les mesures ont d'abord été menées sur des leviers n'ayant subi aucune modification. Puis le dépôt d'une couche métallique ou organique a été apporté sur ces mêmes leviers afin de simuler l'adsorption d'espèce chimique. Une autre série de mesures de fréquences a alors été effectuée dans le but de d'évaluer la sensibilité du capteur.
La procédure consiste à installer le support du (des) levier(s) (cf. Fig. 3-3) dans la tête AFM, puis à calibrer le système de détection par laser (Fig. 3-1) en positionnant le faisceau lumineux à l'apex du levier par un jeu de miroirs (Fig. 3-2 et 3-4), le diriger au centre du quadrant de photodiodes de détection (Fig. 3-5).

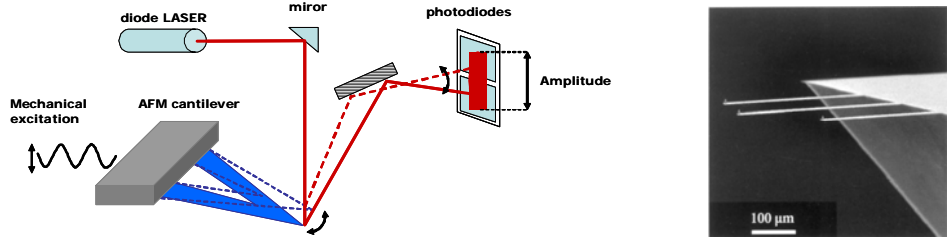

**Natural frequencies of free rectangular cantilevers**
The fundamental vibration mode of a cantilever, i.e. the one with the lowest resonance frequency, always corresponds to a flexural vibration mode. The natural frequencies related to this vibration mode will be computed in the following sections.

**2.1 Analytical calculation**

For many practical applications, the natural frequencies of uniform beam can be determined with sufficient accuracy by an analytical approach (Stokey, 1988). The classical approach is based on the use of the relation which relate the curvature of the beam to the bending moment at each section of the beam :

$$M = EI\chi = EI\frac{\partial^2 y}{\partial x^2}$$

This equation is based upon the assumptions that the material is homogeneous, isotropic and obeys Hooke's law. The beam must be straight, and with a uniform cross section. In order to neglect the shear deflection, the lateral displacement in direction of the Y axis y(x,t) have to be small and the beam have to be long compared to cross sectional dimensions.
E modulus of elasticity in tension and compression (Young's modulus), I area moment of inertie , mass density, area of section
The equation of motion for lateral vibration is found by considering the force action on the element which is formed by passing two parallel planes A and B through the beam normal to the longitudinal axis (figure). The vertical elastic shear force acting on section A and B are V and $V' = V + (\partial V / \partial x)dx$ respectively. The sum of the vertical forces acting on the element must equal the product of the mass of the element and the acceleration $(\partial^2 y / \partial t^2)$ in the lateral direction:

$$\frac{\partial V}{\partial x} + \rho S \frac{\partial^2 y}{\partial t^2} = 0 \quad (1) \text{ et } (2)$$

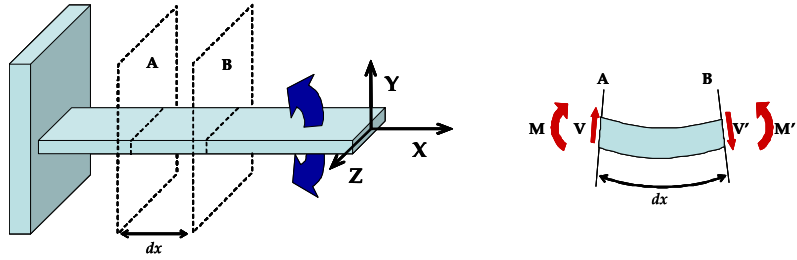

Figure : clamped-free beam executing lateral vibration, (B) element of beam showing shear force and bending moments.

In this element, the equilibrium of the moment drives to $V = \partial M / \partial x$. Other differentials of higher order can be neglected. Then, from eq. 1 and 2, the basic equation for the lateral vibration of beams is given :

$$\frac{\partial^2}{\partial x^2}\left(EI\frac{\partial^2 y}{\partial x^2}\right) + \rho S\frac{\partial^2 y}{\partial t^2} = 0$$

The solution of this equation if EI is constant, is of the form $y = A(x) \times \left[\cos(\omega_n t + \varphi)\right]$ in which A is a function of x only. Substituting $\beta^4 = \rho S \omega^2 / EI$ and dividing eq (3) by $\cos(\omega_n t + \varphi)$ :

$$\frac{d^4 A(x)}{dx^4} = \beta^4 A(x) \quad (3)$$

The following function represents the solution of the eq. (3). The constants A, B, C, D are found from the boundary conditions.

$$A(x) = A \operatorname{ch}\beta x + B \operatorname{sh}\beta x + C \cos\beta x + D \sin\beta x \quad (4)$$

In applying the end conditions in the case of a rectangular beam, clamped at one extremity and free at the other, we have to solve the following equation :

$$1 + \operatorname{ch}\beta\ell \cos\beta\ell = 0 \quad (5)$$

With $\ell$ the lenth of the beam. The solutions of that transcendental equation can be estimated from a numerical approach. As an exemple, with the Newton'method implemented with Matlab2007, we are able to found $\lambda_n = \beta_n \ell$ :

$\lambda_1 = 1.8751$ ; $\lambda_2 = 4.6941$ ; $\lambda_3 = 7.8547$ ; $\lambda_4 = 10.9955$ ; $\lambda_n \approx (2n-1)\frac{\pi}{2} \to \forall n > 4$

The corresponding undamped natural frequency-angular are found by substituting the length of the beamto find each $\beta$ and then :

$$\omega_n = \beta_n^2 \sqrt{\frac{EI}{\rho S}} \quad (6)$$

In the case of rectangular cantilever (one end clamped and the other one free), formula (1) gives the natural frequencies corresponding to the normal flexural modes of vibration of the beam.

$$f_n = \alpha_n \frac{e}{\ell^2} \sqrt{\frac{E}{\rho}} \qquad (7)$$

with $\alpha_n = \frac{1}{4\pi\sqrt{3}} \lambda_n^2$ and e the thickness of the beam.

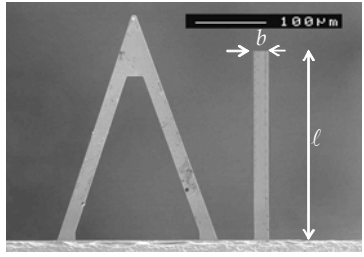

Fig. 1. Description of the main geometrical parameters used in the simulation

## 2.2 Simulation with Finite element method

The purpose of this analysis is to find the transient response from a harmonic load with an excitation frequency in the range 10-1000 KHz, which is near the first Eigen frequencies found with the analytical and experimental approaches. Finite element method (FEM) to numerically solve mechanical equations was chosen. Experimentally measured material constants for metals and silicon at ambient temperature that can be found in the literature were used.

Adaptive meshing and re-meshing is used to tackle the problem of differing length scales. The 3D model makes use of an extruded triangular mesh. High-resolution solutions with minimum computation requirements are tracked. The mesh results in around 250 elements and 6000 degrees of freedom. The numerical model was defined, solved and analyzed in MATLAB using the scripting functions of COMSOL multiphysics (v3.4 with MEMS module) in conjunction with other in-house functions.

The frequency response analysis solves for the steady-state response from harmonic excitation loads. The loads can have amplitudes and phase shifts that depend on the excitation frequency f.

A plate is a thin planar structure, its thickness being less than one tenth of its width. The forces are applied in the direction normal to the plate; the main deformation takes place in the out-of-plane direction. In thin plate theory the transverse shear deformation is neglected.

The default element type is quadratic Lagrange elements. They use second-order polynomials, which is often a good trade off between memory usage and accuracy.

In table 1, we have reported the natural frequencies obtained on a rectangular micro-cantilever. We have compared the results calculated with the formula (1), and those simulated by FEM (for the Mindlin model and for the 3D model). As expected, a very good agreement is obtained between the three different approaches.

| n | Mode | Analytic (KHz) | Mindlin model | FEM 3D |
|---|---|---|---|---|
| 1 | $Long_x$ 1 | 16.4 | 16.5 | 16.5 |
| 2 | $Long_x$ 2 | 103,0 | 103.5 | 103.6 |
| 3 | $Long_x$ 3 | 288.5 | 289.9 | 290.1 |
| 4 | $Long_y$ 1 | --- | --- | 298.5 |
| 5 | Torsional vib. | --- | 531.9 | 531.2 |
| 6 | $Long_x$ 4 | 565.3 | 568.6 | 568.8 |
| 7 | $Long_x$ 5 | 934.5 | 940.9 | 941.2 |

Table 1. Assignment of the first seven natural modes for a rectangular cantilever ($\ell$ = 170 µm, $b$ = 10 µm, thickness $e$ = 0,550 µm, Young's modulus: 143 GPa)

The figure 2 shows the deflection of the beams for the modes 2 and 3.

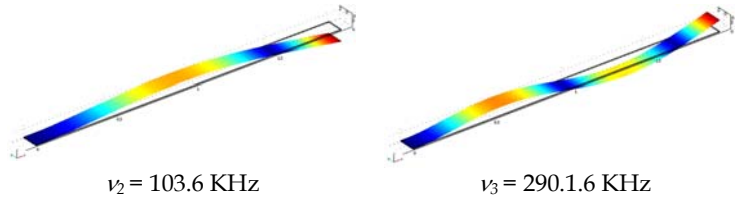

$\nu_2$ = 103.6 KHz            $\nu_3$ = 290.1.6 KHz

Fig. 2. Deformation shapes and corresponding frequencies for two natural vibration modes: n=2 and 3(3D model). The boundaries display the total displacement; the black edges indicate the non-deformed beam geometry; and the white edges delineate the deformed geometry.

**2.3 Experimental and modeling results on a rectangular beam**

By using the classical detection system of an atomic force microscope, we have measured the natural frequencies of a micro-cantilever, which is generally used in intermittent contact mode. A typical spectrum is presented in figure 3. We have also reported on the graph the simulated results (blue line). For the simulation, the thickness of the beam is constituted of 430 nm of silicon and a thin film of gold (70 nm). In order to precisely mesh the structure, we have twice extruded the 2D geometry.

A good agreement is observed between experimental and calculated frequencies. However, it is interesting to note that only one parameter was adjusted in this simulation: the thickness of the thin film of gold. Indeed, an accurate knowledge of this parameter is very difficult to obtain experimentally.

This first part concerning the rectangular cantilevers is very important because it shows the ability of the COMSOL software to precisely calculate the frequencies of very thin films. Moreover, this fact is confirmed by comparing the classical analytical solutions with experimental measurements on bi-layers beams.

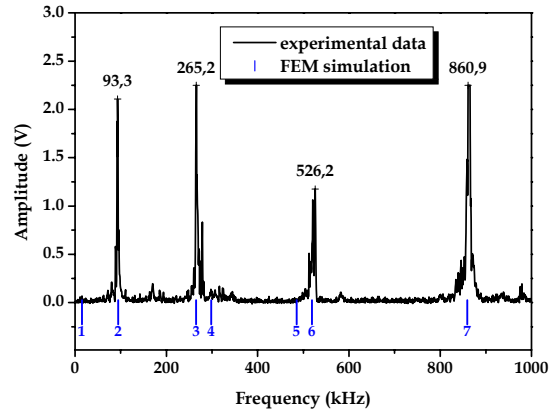

Fig. 3. Experimental spectrum measured on a rectangular beam with the AFM set-up (Veeco Inst.). Underneath, we reported the FEM 3D simulated frequencies (blue line) (épaisseur 50,2 nm, E = 143 GPa,_density = 5000 kg/m$^3$, mesh : de 1024 extruded square elements).

## 2.2 "V" shaped micro-cantilever

In the case of "V" shaped cantilevers, no exact solution (analytic resolution) is known. As an alternative approach, FEM method is particularly useful to calculate the fundamental natural frequencies. In figure 4, we present two SEM images of two commercial AFM cantilevers. The left image correspond to the STS E type and the right to the STS F type.

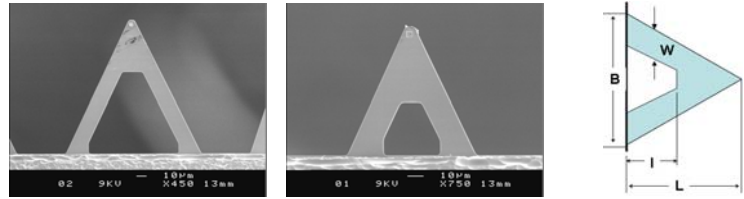

Fig. 4. Examples of commercial AFM-cantilevers and definition of the main geometrical parameters used in FEM simulation.

The main geometrical parameters have been measured using SEM microscope. These parameters are defined in figure 45 and numerical values are reported in table 2. In the same way, experimental and calculated frequencies are reported in table 2.

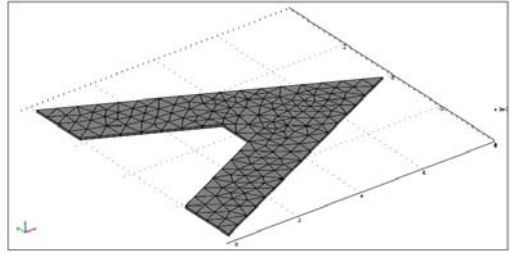

Fig. 6. The 3D mesh produced by extruding a 2D triangular mesh, of very thin "V" shaped cantilever (STS F type).

A particular attention has been devoted to mesh the geometry of our cantilevers. Indeed, it is well-known that very thin structures are very difficult to simulate by FEM. Thank to the mesh module of COMSOL, it is possible to extrude the meshes with a very thin thickness (figure 6).

As an example, the computed deformation of the first fundamental mode of the STS F type cantilever is presented in figure 7.

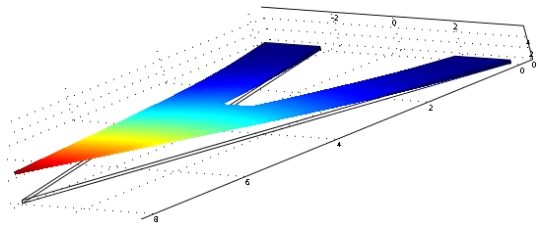

Fig. 7. Deformed shape corresponding to the first vibration mode (STS F type cantilever).

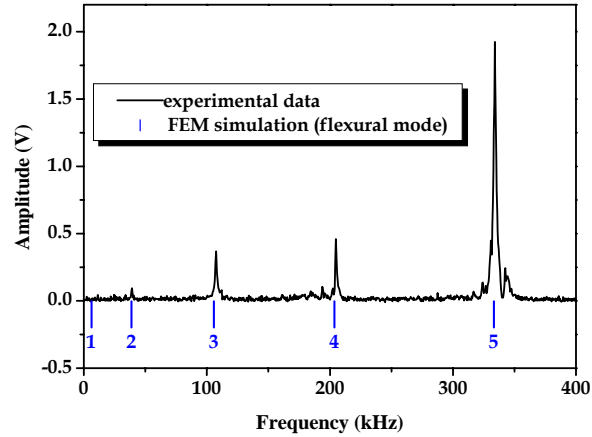

Fig. 8. Experimental spectrum measured on a "V-shaped" micro-cantilever (STS C type). Underneath, the first five FEM calculated frequencies are reported

| Cantilever type | L (µm) | l (µm) | w (µm) | B (µm) | Thickness (**nm**) | Mode n | Experimental frequency (KHz) | Simulated frequency (KHz) | Relative error (%) |
|---|---|---|---|---|---|---|---|---|---|
| STS C | 323 | 235 | 21 | 227 | 680 | 3 | 107.4 | 105.6 | 1.6 % |
|  |  |  |  |  |  | 4 | 204.6 | 203.6 | 0.6 % |
|  |  |  |  |  |  | 5 | 334.0 | 333.3 | 0.2 % |
| STS D | 225 | 133 | 21 | 158 | 700 | 2 | 88.7 | 87.7 | 1.1 % |
|  |  |  |  |  |  | 3 | 233.2 | 234.4 | 0.5 % |
| STS E | 128 | 78 | 18 | 140 | 560 | 2 | 193.4 | 194.7 | 0.8 % |
|  |  |  |  |  |  | 3 | 505.5 | 503.0 | 0.5 % |

Table 2. Measured and simulated resonance frequencies of different "V-shaped" cantilevers (Young modulus: 143 GPa, density = 4800 kg/m$^3$).

A comparison between the natural frequencies measurements and the simulated frequencies is showed in figure 8. We have to point out that the amplitude of the first vibration mode is often very low with our experimental set-up. This amplitude is proportional to the slope of the free end of the beam but almost independent of the lateral deflection. In the simulations, the thickness of the micro-cantilever was the only adjusted parameter.

## 1.1. Modification de leviers par ajout de matière

Silver films have been deposited on the bare core region (called sensitive area) by vacuum thermal evaporation in a glass bell-jar evaporator (Alcatel). The reactor is pumped below 5×10⁻⁶ mbar by oil diffusion and rotary pumps. Silver (Goodfellow 99.99% purity) has been evaporated from a molybdenum crucible.

N,N′-diphenyl-1,4-phenylenediamine (B3) was purchased from Aldrich. Thin films of B3 of controlled thicknesses were deposited by heating, under a vacuum better than 5.10-5 mbar.

An original arrangement of plasma reactor was especially developed for tip applications. It is used to remove dust particles prior to the metallization. Besides, the same plasma device is used to deposit an ultra thin metallic film on the cantilever by plasma sputtering without breaking the vacuum. The metal used here was platinium.

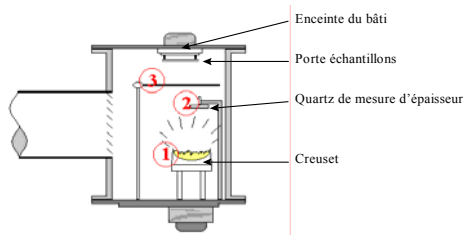          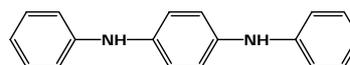

**Figure 1 : Schéma de principe du bâti d'évaporation (équipement Alcatel)**     **Figure 2 : Molécule de B3**

## 3. Conclusion and prospects

In order to evaluate the sensitivity of these cantilevers as chemical transducers, we have started an experimental study. We have deposited by thermal evaporation under vacuum various materials with different thicknesses, from 5 to 20 nm. By taking into account the mass density and the deposited volume, a crude calculation indicates a mass variation varying between 10 and 700 pg. An example of spectra, before and after platinum deposition, is presented in figure 9.

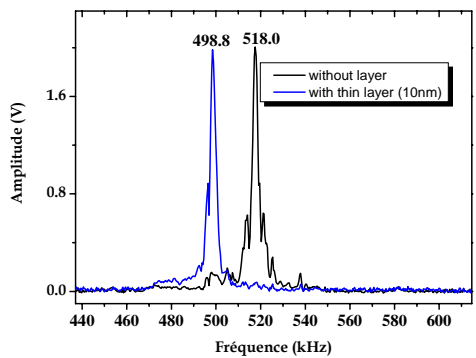

Fig. 9. Spectra of frequency of uncoated and coated micro-cantilever.

A thin solid film of 10 nm of platinum has been deposited onto a STS C type micro-cantilever. A significant frequency shift is observed for the coated cantilever (498.8 kHz) with respect to the frequency of the uncoated cantilever (518 kHz).

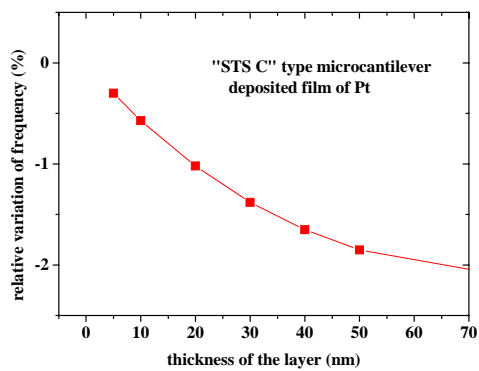

**Fig.10**. Relative variation of the resonance frequency as a function of the deposited layer thickness.

In figure 10, we have simulated the frequency shift as a function of the deposited layer thickness (which can easily convert into added mass). The slope of this curve gives us the sensitivity of the modified cantilever. Such simulations allow us to predict the sensitivity for different geometrical dimensions.

## 7. Acknowledgments
The authors would like to thank M. Collet for useful discussions.